# A Comprehensive Characterization of the Neutron Fields Produced by the Apollon Petawatt Laser

R. Lelièvre[1,2], W. Yao[1,3], T. Waltenspiel[1], I. Cohen[4], A. Beck[5], E. Cohen[5], D. Michaeli[5], I. Pomerantz[4], D. C. Gautier[6], F. Trompier[7], Q. Ducasse[2], P. Koseoglou[8], P.-A. Söderström[9], F. Mathieu[1], A. Allaoua[2] and J. Fuchs[1].

1) LULI - CNRS, CEA, Sorbonne Université, Ecole Polytechnique, Institut Polytechnique de Paris, F-91128 Palaiseau Cedex, France.
2) Laboratoire de micro-irradiation, de métrologie et de dosimétrie des neutrons, PSE-Santé/SDOS, IRSN, 13115 Saint-Paul-Lez-Durance, France.
3) Sorbonne Université, Observatoire de Paris, Université PSL, CNRS, LERMA, F-75005, Paris, France.
4) The School of Physics and Astronomy, Tel Aviv University, Tel-Aviv, 6997801, Israel.
5) Physics department, NRCN, PO Box 9001, Beer-Sheva, Israel.
6) LANL, PO Box 1663, Los Alamos, New Mexico 87545, USA.
7) Laboratoire de dosimétrie des rayonnements ionisants, PSE-Santé/SDOS, IRSN, 92262 Fontenay-aux-Roses, France.
8) Institute for Nuclear Physics, Technische Universität Darmstadt, Schlossgartenstr. 9, 64289 Darmstadt, Germany.
9) Extreme Light Infrastructure-Nuclear Physics (ELI-NP)/Horia Hulubei National Institute for Physics and Nuclear Engineering (IFIN-HH), Str. Reactorului 30, 077125 Bucharest-Măgurele, Romania.

## Abstract

Since two decades, laser-driven neutron emissions are studied as they represent a complementary source to conventional neutron sources, with further more different characteristics (i.e. shorter bunch duration and higher number of neutrons per bunch). We report here a global, thorough characterization of the neutron fields produced at the Apollon laser facility using the secondary laser beam (F2). A Double Plasma Mirror (DPM) was used to improve the temporal contrast of the laser which delivers pulses of 24 fs duration, a mean on-target energy of ~10 J and up to 1 shot/min. The interaction of the laser with thin targets (few tens or hundreds of nm) in ultra-high conditions produced enhanced proton beams (up to 35 MeV), which were then used to generate neutrons via the pitcher-catcher technique. The characterization of these neutron emissions is presented, with results obtained from both simulations and measurements using several diagnostics (activation samples, bubble detectors and Time-of-Flight detectors), leading to a neutron yield of ~$4\times10^7$ neutrons/shot. Similar neutron emissions were observed during shots with and without DPM, while fewer X-rays are produced when the DPM is used, making this tool interesting to adjust the neutrons/X-rays ratio for some applications like combined neutron/X-ray radiography.

## I. INTRODUCTION

Numerous experiments have demonstrated the capacity of ultra-intense lasers to generate beams of particles, including neutrons of MeV range [1,2,3,4]. These processes are initiated by the interaction of an intense laser with a thin target (~ µm) which first accelerates electrons. These can then accelerate ions (mainly protons) through a variety of mechanisms [5], among which the robust Target Normal Sheath Acceleration (TNSA) mechanism [6,7,8], which generates a typical exponentially decreasing proton energy distribution. Meanwhile, a large amount of X-rays are produced within the target, mostly due to Bremsstrahlung emissions [9,10,11,12].

The production of secondary neutrons can take place through several mechanisms, among which: the pitcher-catcher technique [2,13,14,15,16], beam-fusion reactions [17,18] and photoneutron generation [19,20]. The pitcher-catcher technique has been widely used in many experiments, demonstrating its ability to produce intense neutron emissions. This technique uses a double target system: a first target (the pitcher) is irradiated by the laser to accelerate ions, which are then intercepted by a second target (called catcher or converter), inducing within it nuclear reactions that produce neutrons. Low-Z converters (LiF or Be) are usually used to take advantage of the interesting cross sections of (p,n) reactions in these materials to produce neutrons from the significant emissions of low energy protons (few MeV) produced by TNSA.

Laser facilities, as new neutron sources, are mainly characterized by their compact size and the possibility to produce significant neutron emissions, from sub-MeV to tens of MeV, in very small intervals (< ns) leading to high-brightness neutron pulses [21,22] with reduced radiological constraints compared to conventional sources (no fission products and low material activation) [23,24]. Thus, laser-driven neutron sources seem particularly interesting for several applications: radiography of light materials (such as lithium batteries) [25,26], contraband detection (explosives, narcotics) [27,28,29], astrophysics [30,31] or material analysis using neutron resonance spectroscopy [32,33]. These laser facilities therefore represent as many potential new neutron sources, in addition to reactors or accelerator-based spallation sources, the number of which being today insufficient, since demand is struggling to be satisfied due to high construction costs and significant radiological constraints of conventional sources [4].

We present here the first quantitative measurements of neutron emissions carried out with the Apollon petawatt laser. These neutrons were produced by the interaction of protons, accelerated from Al or Si targets, with a LiF converter. The neutron emissions were characterized by several diagnostics, including activation samples, bubble detectors [34] and ultra-fast organic scintillators as neutron Time-of-Flight (nToF) detectors. A Double Plasma Mirror (DPM) [35,36,37] was also used during some shots to improve the laser/target interaction and adjust the proton acceleration as well as the secondary X-ray emissions and neutron production.

Section II gives an overview of the experimental setup and the proton acceleration, with and without DPM. Section III describes the simulations performed with the Monte-Carlo transport code Geant4 [38] to predict the neutron emissions and the response of some of the detectors used. Section IV shows the experimental measurements compared to the simulations and some comparisons between the results obtained with and without the use of the DPM. Finally, Section V provides a summary of the main results, a conclusion of this work and some prospects.

## II. EXPERIMENTAL SETUP

The experiment was performed using the secondary laser beam (F2) of the Apollon laser facility in Saclay, France [39]. This beam uses a Ti:Sapphire laser which delivers pulses of 24 fs duration, a mean on-target energy of 10.9 J and a central laser wavelength of 815 nm, spanning from 750 to 880 nm. The 140 mm diameter beam was focused using an f/3 off-axis parabola (OAP), inducing an elliptical focal spot (2.8×3.7 µm FWHM) which contains 42% of the total laser energy, resulting in an on-target peak intensity of around $2\times10^{21}$ W/cm².

The laser was used with its inherent temporal contrast but also with a Double Plasma Mirror (DPM), placed inside the chamber between the OAP and the target (see Fig. 1). A plasma mirror is composed of a polished glass slab through which prepulses are first transmitted. Then, when the main pulse arrives, the intensity increases and ionization occurs on the plasma mirror surface. As the electron density exceeds the critical density, it allows to reflect most of the main pulse. The DPM induces a lower mean on-target energy (~5.7 J), due to its 52% reflectivity, but it improves the laser contrast by reducing the pre-pulses. This avoids heating and ionizing the target before the arrival of the main pulse and allows shooting on thinner targets thanks to the improvement of the laser/target interaction.

### A. Proton acceleration
The beam was focused on targets with an 45° angle of incidence. Different targets were used: 0.8 µm, 1.5 µm and 2 µm thick aluminum targets for direct shots (without DPM), and thinner silicon targets, from 20 nm to 300 nm, for shots with the DPM. The produced protons were accelerated from the rear surface of the target using the TNSA mechanism, with energies up to 25.3 MeV in direct shots and 35.2 MeV during DPM shots.

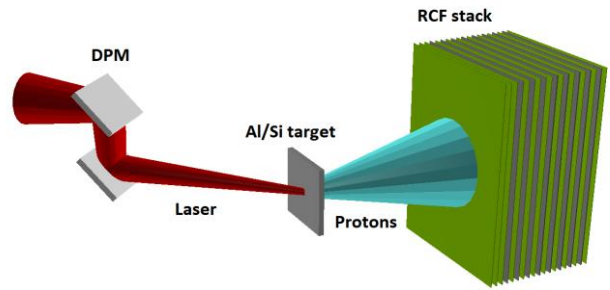

**Fig. 1.** Schematic view of the experimental setup.

As shown in Fig. 1, the proton spectra were measured using stacks of EBT3 Gafchromic RadioChromic Films (RCF) [40] and aluminum filters, placed at 25 mm from the target. A deconvolution can be done from the doses obtained on each RCF films to retrieve the proton spectra. Fig. 2 shows two proton spectra, produced during direct and DPM shots.

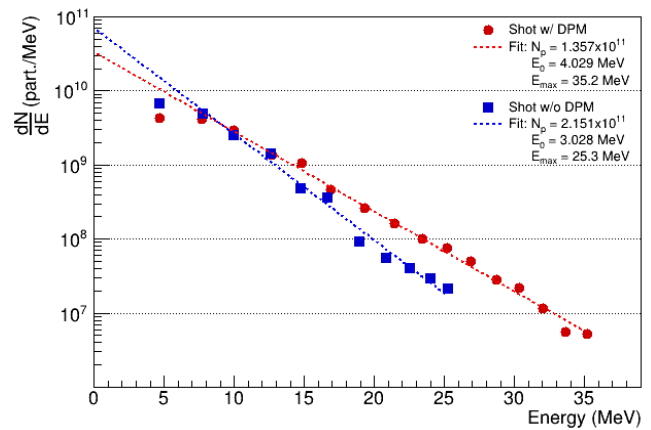

**Fig. 2.** Measured proton spectra, extracted from RCF stacks for a shot with DPM on a 200 nm Si + 50 nm Al target (red circles), and a shot without DPM on a 1.5 µm Al target (blue squares). Respective exponential fits are represented by dashed lines.

As shown in Fig. 2, the proton spectra are well-fit using exponential functions with:

$$\frac{dN}{dE}(E \leq E_{max}) = \frac{N_p}{E_0}\exp\left(-\frac{E}{E_0}\right) \quad (1)$$

Where $N_p$ is the total proton number, $E_0$ is the slope of the spectrum and $E$ is the proton energy. The first RCF films were not considered for the fit function calculation because the proton divergence was higher at low energies and a large part of the proton beam was not covered by those, so the values obtained with these films are certainly underestimated.

Although the DPM reduces the on-target energy by almost a factor 2, the experimental measurements demonstrate the possibility to obtain higher maximum proton energies while the total proton numbers is of the same order of magnitude, compared to the results obtained during shots without DPM. This means that the DPM has an important role in the proton acceleration enhancement by improving the laser/target interaction, leading to better conversion efficiencies. The detailed investigation of these interactions will be the focus of a separate paper.

### B. Neutron emission

#### 1. Converter

To produce neutrons, the RCF stack was replaced by a LiF converter, placed 15 mm behind the TNSA target (see Fig. 4). A low-Z converter was preferred to take advantage of good cross sections of (p,n) reactions for low proton energies (< 10 MeV, see Figure 3), corresponding to more than 90% of the protons we generate.

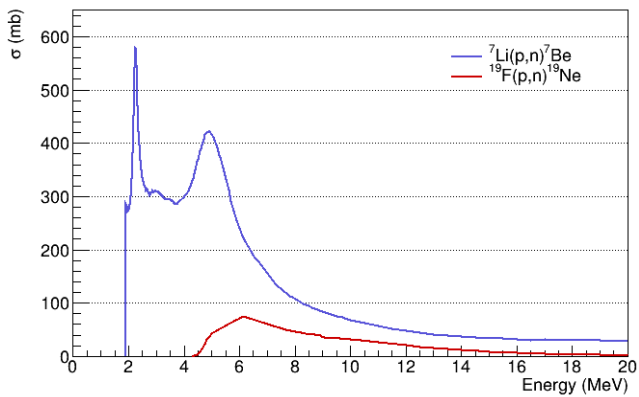

**Fig. 3.** Cross sections of the $^{7}$Li(p,n)$^{7}$Be and the $^{19}$F(p,n)$^{19}$Ne reactions, as given by the ENDF/B-VIII.0 [41] and TENDL-2019 [42] libraries, respectively.

The converter was 4 mm thick, which ensures that no proton with energy below ~30 MeV can pass through it. Most neutrons were produced by the $^{7}$Li(p,n)$^{7}$Be reaction; another reaction can occur ($^{19}$F(p,n)$^{19}$Ne) but produces much fewer neutrons due to lower cross sections and a higher threshold energy (as can be seen in Fig. 3).

#### 2. Diagnostics

The neutron emissions were characterized using bubble detectors, scintillators used as neutron Time-of-Flight (nToF) detectors, activation samples and direct measurement of the total number of neutrons produced by measuring the activity of the $^{7}$Be residual nuclei inside the LiF converter.

As shown in Fig. 4, three different activation samples were used (copper, indium and magnesium), in which neutrons induce nuclear reactions and generate radionuclides that are then measured using gamma spectrometry. Knowing the quantity of the radionuclides produced, the cross sections and the threshold energies of these reactions, information about the energy distribution of neutrons can be obtained and activation spectrometry can be performed when several samples, sensitive to different neutron energies, are used.

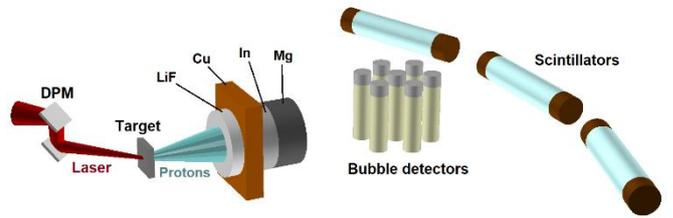

**Fig. 4.** Schematic view of the setup and diagnostics used during neutron generation with the DPM.

For the nToF measurements, several ultra-fast organic scintillators were placed at different angles and distances compared to the LiF converter to study the angular and energy dependence of the neutron emissions (see Fig. 5). The detectors are composed of an 1" diameter and 40 cm long scintillator tube (EJ-254) [43] connected to one photomultiplier tube (9112B) [44] on each side to collect the scintillation photons [45]. The signal was then digitized for a duration of 1 ms at a sampling frequency of 500MS/s using a CAEN VX1730B digitizer. To avoid saturation effects that can be induced by the strong X-ray emissions, these scintillators were shielded with 15 cm thick lead bricks on the front face and 5 cm thick lead bricks on the top, bottom and rear faces.

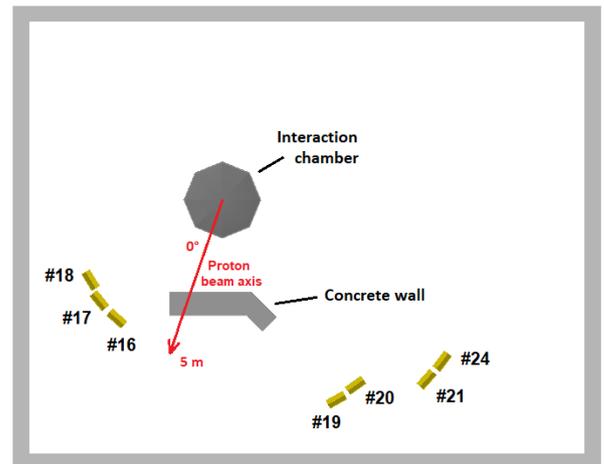

**Fig. 5.** Positions of nToF detectors (in yellow) inside the experimental room. The target normal axis corresponding also to the main axis of proton emission is represented by the red arrow.

## III. MONTE-CARLO MODELING

### A. Neutron production

Simulations were carried out, using the Monte-Carlo transport code Geant4 [38], to estimate the total number of neutrons and the angular distribution of these emissions produced by the interaction of the proton beams, we experimentally characterized, with the LiF converter.

These simulations used cross sections coming from the ENDF/B-VIII.0 [41] and TENDL-2019 [42] libraries to reproduce all the nuclear reactions inside the converter. Averaged proton spectra over 13 and 8 shots, obtained respectively from direct and DPM shots, were used to simulate the neutron emissions, in order to take into account the shot-to-shot variability of the laser/target interaction and the proton beam.

These protons were injected into a simulated 4 mm thick LiF converter using conical beams with a half angle of 15°, which approximates the mean divergence of the typical TNSA proton beam [46]. Virtual cylindrical detectors were placed from 0° to 180° (in 22.5° steps) to get the neutron fluence at different angles. Another virtual spherical detector covering the 4π sr solid angle around the converter was added to obtain the total number of neutrons emitted. The aluminum chamber and the concrete walls (see Fig. 5) were also considered in the simulations to get closer to the real conditions of the experiment.

Fig. 6 shows the simulated neutron fluences at different angles, considering direct and DPM shots. The neutron emissions are very similar and, in both cases, almost isotropic. A dip can be observed at 90°, this is due to the diameter of the LiF converter (22 mm) which is greater than its thickness, so the neutrons emitted transversally need to pass through more material to exit the converter, leading to more deflection or absorption of neutrons.

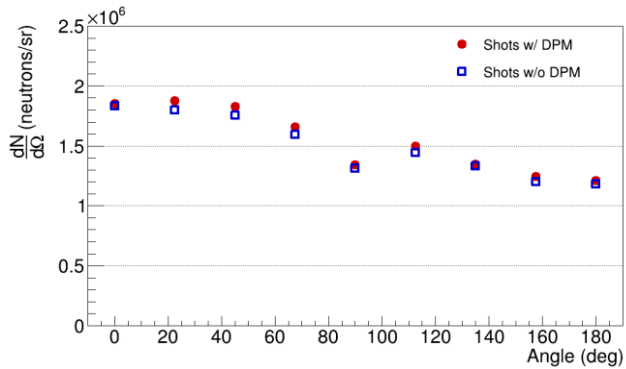

**Fig. 6.** Neutron fluences at different angles obtained from the Geant4 simulations for direct and DPM shots.

Fig. 7 shows the simulated energy differential neutron spectra obtained in the forward direction (rear surface of the converter) and the backward direction. We see that the most energetic neutrons are preferentially emitted forward, which agrees with the kinematics of the interactions.

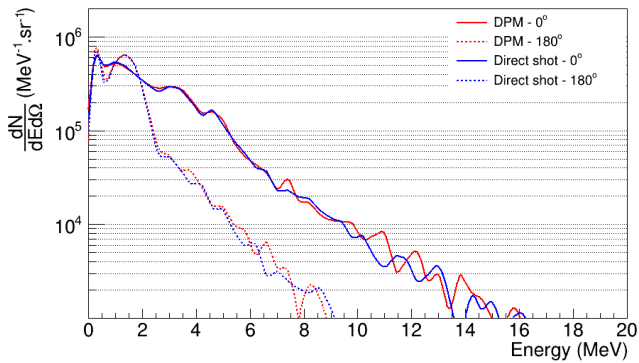

**Fig. 7.** Simulated energy differential neutron spectra in forward (0°) and backward (180°) directions for shots with and without DPM.

In the subsequent sections, the neutron spectra in the forward direction will be used to predict the response of the activation diagnostic and the scintillators. The total number of neutrons computed by these simulations was $2.694 \times 10^7$ neutrons/shot in the direct shot configuration and $2.781 \times 10^7$ neutrons/shot when the DPM is used. As shown in Fig. 7, the simulated neutron spectra obtained with and without the DPM are very similar, so the presence of the DPM does not seem to affect the neutron emissions.

Finally, additional simulations were made to determine the number of neutrons produced exclusively from the Li nuclei. These numbers were found to be $2.521 \times 10^7$ neutrons for direct shots and $2.601 \times 10^7$ neutrons for shots with DPM, leading to $^7$Be activities of 3.8 Bq and 3.9 Bq, respectively. Thus, the neutrons are mainly produced from the Li nuclei which contribute to around 93.5% of the neutron emissions.

### B. Design of the activation diagnostic

Several criteria were used to select the appropriate activation samples. The neutron-induced radionuclides must emit gamma-rays with good branching ratios and suitable half-lives, typically of several hours or days, which is long enough to avoid the complete decay before the measurement but not too long to reduce the measurement time in gamma spectrometry. The chosen elements have also to be chemically stable and the isotope with which the reaction of interest occurs must have a large relative abundance to avoid confusions between neutron reactions produced by the different isotopes in the same sample. Finally, the reactions of interest must cover different energy ranges to get information on the energy distribution of neutrons.

Considering all these criteria, several elements were selected to compose 3 layers (see Table I), corresponding to different reactions and neutron energy ranges, due to the energy dependence of their cross sections. We used the simulated neutron spectra shown in Fig. 7 to calculate the number of produced nuclei $N_0$ and activities $A_0$ inside all these elements to select the most suitable ones for our purpose. A conical beam of $10^7$ neutrons and samples of 1" diameter and 10 mm thick were considered in these simulations, which is not necessarily representative of the real dimensions of the samples we ultimately employed, but facilitates comparisons across various materials.

| Element | Reaction | Half-life | Threshold (MeV) | $N_0$ | $A_0$ (Bq) |
|---|---|---|---|---|---|
| Au | $^{197}$Au(n,g)$^{198}$Au | 6.17 d | 0 | $7.3 \times 10^4$ | 0.22 |
| Cd | $^{114}$Cd(n,g)$^{115}$Cd | 2.23 d | 0 | $5.9 \times 10^3$ | 0.02 |
| Cu | $^{63}$Cu(n,g)$^{64}$Cu | 12.7 h | 0 | $8.5 \times 10^3$ | 0.13 |
| Mn | $^{55}$Mn(n,g)$^{56}$Mn | 2.58 h | 0 | $4.4 \times 10^3$ | 0.33 |
| W | $^{186}$W(n,g)$^{187}$W | 24 h | 0 | $8.9 \times 10^3$ | 0.07 |
| Zn | $^{64}$Zn(n,p)$^{64}$Cu | 12.7 h | 0.5 | $5.5 \times 10^3$ | 0.08 |
| In | $^{115}$In(n,n')$^{115m}$In | 4.49 h | 0.5 | $3.2 \times 10^4$ | 1.37 |
| Mg | $^{24}$Mg(n,p)$^{24}$Na | 15 h | 5 | $4.7 \times 10^2$ | 0.01 |
| Al | $^{27}$Al(n,a)$^{24}$Na | 15 h | 4.5 | $4.1 \times 10^2$ | 0.01 |
| Fe | $^{56}$Fe(n,p)$^{56}$Mn | 2.58 h | 3 | $6.0 \times 10^2$ | 0.04 |

**Table I.** List of potential elements considered as activation samples and that were assessed here.

These estimations show that the best element for the first layer was gold or manganese. But due to the difficulty of acquiring a gold sample with such dimensions and considering the short half-life of the radionuclide produced in the manganese sample, these two elements were excluded. Thus, the copper sample appears to be a good compromise, both in terms of number of nuclei produced, activity and half-life. Then, the indium sample was chosen for the second layer, because of inducing much more activation compared to the zinc sample. And the third layer will be composed of the magnesium sample, because it produces more activation compared to the aluminum sample and has a radionuclide with a longer half-life than the one induced by the iron sample.

Finally, a modeling of a NaI spectrometer used for the gamma spectrometry measurement was made using Geant4, to optimize the thickness of the samples. This is necessary because a thick sample absorbs more neutrons and produces more activation but acts also as a shield for the following sample and increases the self-absorption of gamma-rays, thereby reducing the counting rate during the gamma spectrometry measurements. Different thicknesses for the three samples, from 3 to 30 mm, were considered in the simulations. This resulted in an optimized thickness of 6 mm for the first and the second layer, and 10 mm for the third layer.

In summary, the elements selected for the activation diagnostic were a copper slab (thickness = 6 mm, dimensions = 24×32 mm², ρ = 8.96 g/cm³), a cylindrical indium sample (thickness = 6 mm, diameter = 22 mm, ρ = 7.30 g/cm³) and a cylindrical magnesium sample (thickness = 10 mm, diameter = 22 mm, ρ = 1.74 g/cm³). This activation stack was placed at 0°, directly behind the LiF converter (as shown in Fig. 4) to capture as many neutrons as possible and to maximize the activation of the samples. Additional simulations were carried out considering the simulated neutron spectra presented in Fig. 7, the dimensions and the position of these activation samples, to calculate simulated activities which can be compared to those measured (see Section IV.A.2).

### C. Scintillation modeling

Geant4 was also used to simulate the nToF signal of the scintillators using, as inputs, the simulated neutron spectra obtained at different angles thanks to the calculations described in Section III.A. The concrete walls, the aluminum chamber and the lead shielding were considered to take into account the influence of neutron scattering on the nToF signals. Scintillation processes were modeled according to a previous energy calibration performed using a $^{137}$Cs source [45], allowing to obtain simulated nToF signals in mV (see Section IV.C).

Other simulations considering monoenergetic neutrons were also performed to determine the response function of the scintillators. Due to the temporal characteristics of these scintillators (rise time, decay time, …), the signal induced by monoenergetic neutrons spans a time interval greater than the corresponding Time-of-Flight (ToF) (see Fig. 8).

A deconvolution procedure will therefore be applied to the measured nToF signals to obtain neutron spectra by subtracting the contribution of neutrons of a certain energy to higher ToF and lower energies (for more detail see Section IV.C).

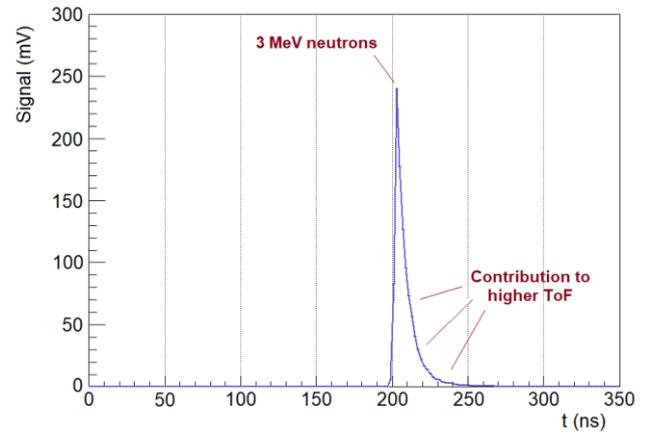

**Fig. 8.** Simulated nToF signal obtained for a scintillator placed at 495 cm from a conical source of 3 MeV neutrons with a fluence of $10^5$ neutrons/sr.

## IV. NEUTRON FIELD CHARACTERIZATION

### A. Activation measurements

The activation samples were measured by gamma spectrometry using a 3x3" NaI scintillator shielded by 10 cm of lead and a layer of copper to absorb the fluorescence X-rays emitted by the lead shielding. The signals were digitized with an OSPREY Multi-Channel Analyzer (MCA) and the Genie2000 software was used to analyze the gamma-ray emission spectra by subtracting the background noise and performing peak fitting.

Due to their dimensions, the activation samples cannot be considered as point sources because self-absorption effects of gamma-rays can occur and affect the measurements. So, the efficiency calibrations associated to these different geometries need to be determined and this was done using a dissolution procedure, performed by the Laboratoire National Henri Becquerel (LNHB, CEA), for the indium and magnesium samples.

For the copper sample and the LiF converter, the efficiency calibrations were calculated using the efficiency transfer method [47,48] based on the Moens concept [49], allowing to determine efficiency calibrations for different geometries from an $^{152}$Eu calibration point source and Geant4 simulations.

#### 1. $^7$Be residuals in LiF converter

Two main reactions induce the emission of neutrons from a LiF converter: $^7$Li(p,n)$^7$Be and $^{19}$F(p,n)$^{19}$Ne. So, the measurement of the LiF converter in gamma spectrometry and the determination of the quantity of these residuals give information about the total number of neutrons produced. The $^{19}$Ne radionuclide cannot be measured by gamma spectrometry due to his very short half-life ($T_{1/2}$ = 17.22 s).

However, the $^7$Be radionuclide can be easily measured and, according to the simulations (see Section III.A), the reaction producing this residual is responsible for almost all the neutron emissions. Thus, measuring this residual gives a good estimation of the total production of neutrons.

A LiF converter, on which we cumulated neutrons produced during 20 direct shots as well as 20 DPM shots, was measured in gamma spectrometry, resulting in an activity of $^7$Be of 221.3 ± 12 Bq. This induces that around 3.67×10$^7$ neutrons/shot were produced from the Li nuclei, which contribute to 93.5% of the neutron production (see Section III.A). The total number of neutrons produced during these shots is therefore around 3.93×10$^7$ neutrons/shot. This value is, on average, 1.46 and 1.41 times higher than the simulation predictions for a direct shot and a DPM shot, respectively (as shown in Table II).

|  | A/shot (Bq) | $N_{Be-7}$/shot | $N_{neutrons}$ (neutrons/shot) |
|---|---|---|---|
| *Simulation (Direct shots)* | 3.8 | 2.52×10$^7$ | 2.70×10$^7$ |
| *Simulation (DPM shots)* | 3.9 | 2.60×10$^7$ | 2.78×10$^7$ |
| LiF converter measurement | 5.53 | 3.67×10$^7$ | 3.93×10$^7$ |

**Table II.** Simulated and measured activity, as well as number, $N_{Be-7}$, of $^7$Be nuclei residuals produced during a series of direct and DPM shots. The total neutron production per shot is also shown.

### 2. Activation samples

Activation samples were used over two separate series, one of 20 direct shots and the other of 5 DPM shots. In both cases, we accumulated the activation to increase the probability of obtaining measurable activities. Table III presents the results of the gamma spectrometry measurements of these samples and a comparison with the simulated values.

|  | Reaction | $A_{mes}$/shot (Bq) | ε (%) | $A_{sim}$/shot (Bq) |
|---|---|---|---|---|
| Direct shots | $^{63}$Cu(n,g)$^{64}$Cu | 17.60 | 6.84 | 0.46 |
|  | $^{115}$In(n,n')$^{115m}$In | 0.93 | 13.66 | 1.04 |
|  | $^{24}$Mg(n,p)$^{24}$Na | < DL* | - | 0.005 |
| DPM shots | $^{63}$Cu(n,g)$^{64}$Cu | < DL* | - | 0.56 |
|  | $^{115}$In(n,n')$^{115m}$In | 0.75 | 36.75 | 1.21 |
|  | $^{24}$Mg(n,p)$^{24}$Na | < DL* | - | 0.013 |

*DL: Detection Limit*

**Table III.** Measured activities, $A_{mes}$, after the last shot of the two series, one in direct mode, the other using the DPM, and the corresponding uncertainties, ε. $A_{sim}$ are the simulated activities calculated in Section III.B from the spectra presented in Fig. 7. These simulated activities were normalized by the number of $^7$Be nuclei observed (i.e. 1.46 and 1.41 times, for direct and DPM shots, respectively), to consider the actual production of neutrons, which is greater than expected.

No measurable activation was induced in the magnesium sample, either during direct or DPM shots, meaning that neutrons with energies above 5 MeV are produced in too small number to induce measurable activities, as predicted by the simulations.

The activation of the copper sample is much higher than the simulated value, this could partly be explained by a greater number of low-energy neutrons emitted compared to the predictions from the simulations. But, the copper sample was also not fully covered by the LiF converter and, due to the divergence of the protons, some of them were able to directly interact with the copper sample, resulting in a significant production of $^{64}$Cu via the $^{65}$Cu(p,d)$^{64}$Cu reaction, thus likely explaining the large difference between simulated and measured activities.

The measured activities of the indium samples are on the contrary close to the simulated ones, the simulations seem therefore to give good predictions of neutron emissions in its sensitivity range (see Fig. 9). Moreover, the activities obtained from direct and DPM shots are also close to each other and the overlapping uncertainties lead to similar neutron emissions during these two shot configurations.

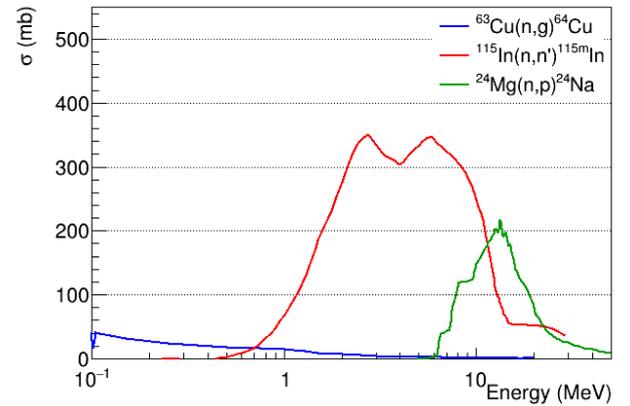

**Fig. 9.** Experimental cross sections of neutron-induced reactions in the samples, according to the EXFOR library [50].

Due to the relatively flat cross section profile of the reaction induced in the indium sample between 2 and 10 MeV (see Fig. 9), we can estimate the approximative number of neutrons emitted in this energy range using the equation:

$$N_\Omega = \frac{A_{mes}}{\lambda \times n_{shot}} \times \frac{1}{\bar{\sigma} \times t \times n \times \chi} \times \frac{1}{\Omega} \quad (2)$$

With:

$N_\Omega$, in neutrons/sr/shot
$A_{mes}$, the measured activity (in Bq)
$\lambda$, the decay constant (s$^{-1}$)
$n_{shot}$, number of shots
$\bar{\sigma}$, the average cross section (m$^{-2}$)
$t$, the sample thickness (m)
$n$, number density (m$^{-3}$)
$\chi$, the isotope abundance
$\Omega$, the solid angle covered by the sample (sr)

Considering an average cross section of 314 mb in the considered energy range and the measured activities presented in Table III, neutron fluences of 1.689×10$^6$ ± 22.1% neutrons/sr/shot were obtained for direct shots and 1.136×10$^6$ ± 45.1% neutrons/sr/shot for DPM shots. Note that these results are averaged over several shots and may therefore be affected by the decay of the $^{115m}$In radionuclide between shots, leading to underestimated values. Thus, a correction was applied to take into account the mean decay time between shots due to the repetition rate of approximately 1 shot every 8 minutes.

### B. Bubble detectors

Two types of bubble detectors, manufactured by Bubble Technology Industries (BTI), were used during this experiment: bubble dosimeters (BD-PND) and a bubble spectrometer (BDS). They were placed inside the interaction chamber, in the target normal axis (0°), at 60 and 20 cm from the converter, respectively. These detectors are made of an elastic polymer containing droplets of superheated liquid that turn into small visible bubbles when neutrons interact with them. Calibrations to obtain the deposited doses from the number of bubbles observed inside the detectors are given by BTI. These calibrations were verified at the CEZANE facility (IRSN, Cadarache), using a $^{252}$Cf neutron source. Significant discrepancies were observed with the given calibrations, with overestimations of the response of a factor of more than 3 for some detectors. Thus, the new calibration values will be considered in the following analyses.

The bubble spectrometer is composed of 6 bubble dosimeters with different sensitivity ranges (from 10 keV to 20 MeV), allowing to obtain an energy distribution of neutrons. The BD-PND are sensitive to neutrons from 100 keV to 20 MeV with a constant energy response; a neutron fluence can therefore be calculated from the measured dose. These BD-PND were employed during the same series of direct and DPM shots as those where the activation samples were used. The results obtained with these BD-PND lead to neutron fluences of $4.377 \times 10^6 \pm 17.4\%$ neutrons/sr/shot for the direct shots and $4.722 \times 10^6 \pm 40.8\%$ neutrons/sr/shot during the DPM shots.

These neutron fluences are much higher than the ones obtained with the indium samples. This could be explained by the larger sensitivity range of the bubble dosimeters, from 100 keV to 20 MeV, compared to only 2 to 10 MeV for the indium samples. The positions and the solid angles covered by these two diagnostics can also explain this difference because the bubble dosimeters were placed at 60 cm from the converter and captured neutrons emitted at 0°, while the indium samples were very close to the converter and covered a solid angle of around 2.3 sr, reducing the average neutron fluence intercepted considering the angular distribution of neutrons, preferentially emitted at 0° (see Fig. 6).

The integral of the expected neutron spectra (shown in Fig. 7) for these two energy ranges was calculated and suggests that we should have more than 2 times more neutrons in the sensitivity range of the bubble dosimeters compared to that of indium samples, which is consistent with the given explanation.

The average energy to which these bubble dosimeters are sensitive can be calculated by determining the probability density function, *p(E)*, considering the response of these detectors and the expected shape of the neutron spectra:

$$p(E) = \frac{1}{Y_{sim}} \times \frac{dY}{dE}(E) \quad (3)$$

With:

$$\frac{dY}{dE}(E) = \eta(E) \times \frac{dN}{dEd\Omega}(E)_{sim} \times \Omega \quad (4)$$

$$Y_{sim} = \int \frac{dY}{dE} dE \quad (5)$$

Where $\eta(E)$ is the number of events per neutron, $\frac{dN}{dEd\Omega}(E)_{sim}$ the simulated neutron spectrum, $\Omega$ the solid angle and $Y_{sim}$ the total number of event (i.e. number of bubbles).

The average energy is then defined as:

$$\bar{E}_{sim} = \int E \times p(E) \, dE \quad (6)$$

And an associated neutron fluence can be calculated with equation (7) and normalized by the measured value $Y_{exp}$ using equation (8).

$$\overline{\frac{dN}{dEd\Omega}}_{sim} = \int \frac{dN}{dEd\Omega}(E)_{sim} \times p(E) \, dE \quad (7)$$

$$\overline{\frac{dN}{dEd\Omega}}_{exp} = \frac{Y_{exp}}{Y_{sim}} \times \overline{\frac{dN}{dEd\Omega}}_{sim} \quad (8)$$

This analysis was also done for the indium samples, considering the number of events per neutron, $\eta(E)$, as:

$$\eta(E) = \sigma(E) \times t \times n \times \chi \quad (9)$$

In this case, the total number of events $Y$ is defined as the number of $^{115m}$In nuclei simulated or measured. The results of these calculations are presented in Table IV and plotted in Fig. 10.

|  | Diagnostic | $\bar{E}_{sim}$ | $\overline{\frac{dN}{dEd\Omega}}_{sim}$ | $\overline{\frac{dN}{dEd\Omega}}_{exp}$ |
|---|---|---|---|---|
| Direct shots | Bubble dosimeters | 2.70 | $5.034 \times 10^5$ | $9.204 \times 10^5$ |
| | Indium sample | 3.36 | $3.913 \times 10^5$ | $3.163 \times 10^5$ |
| DPM shots | Bubble dosimeters | 2.70 | $4.861 \times 10^5$ | $9.817 \times 10^5$ |
| | Indium sample | 3.39 | $3.760 \times 10^5$ | $2.495 \times 10^5$ |

**Table IV.** Average energies, simulated and normalized neutron fluences for the bubble dosimeters and the indium samples used during the series of direct and DPM shots.

Fig. 10 shows a comparison between these neutron fluences, the neutron energy spectrum at 0° measured with the bubble spectrometer and the simulated neutron spectra for a direct and a DPM shot, which have been multiplied, respectively, by a factor 1.46 and 1.41 to normalize them considering the total number of neutrons observed from the $^7$Be measurement.

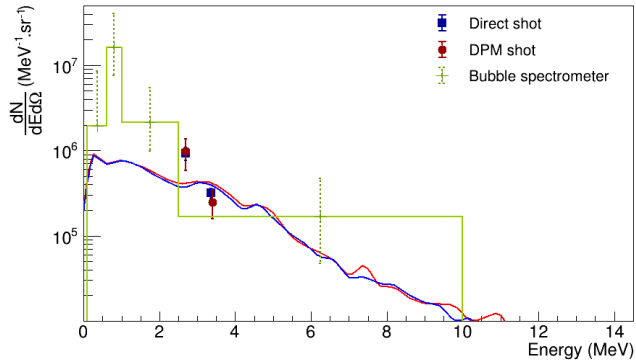

**Fig. 10.** Neutron energy spectrum measured with the bubble spectrometer (green line) during the series of direct shots, simulated neutron spectra for a direct shot (blue line) and a DPM shot (red line) and markers showing the neutron fluences, calculated from the bubble dosimeters and the indium samples, normalized by the measured values.

The neutron fluences obtained during the direct and DPM shots with the bubble dosimeters and the indium samples are very close, which is consistent with the hypothesis of similar neutron productions, demonstrated previously, during these two shot configurations.

The Geant4 simulations seem to overestimate the fluence at the average sensitivity energy of the indium samples and underestimate the fluence at that of the bubble dosimeters, resulting in a probable underestimation of the low-energy part of the simulated neutron spectrum. This can also be observed on the spectrum obtained by the bubble spectrometer, where a difference of an order of magnitude is shown for energies below 1 MeV. But, due to the uncertainties associated to this measurement, the quantification of this difference is not really significant and it could also be partly explained by the detection of neutrons scattered inside the interaction chamber.

### C. nToF measurements

Several organic scintillators, placed at different angles and distances from the converter (see Fig. 5), were used to detect the neutrons using the Time-of-Flight technique. This method relies on the different velocities of the neutrons, depending on their energy, which produce a signal spread over time, allowing to obtain their ToF and so the information about the energy distribution of neutrons. The X-rays also induce a signal in the scintillators and, knowing that they propagate at the speed of light, serves as time reference to define the instant at which the neutrons are emitted. However, since the X-ray emission is intense, an appropriate lead shielding must be installed all around the scintillators to avoid the saturation effects of the photomultipliers tube used to collect the light produced by the scintillators. Limiting the signal induced by the X-rays is also important to improve the detection of the high energy neutrons, whose nToF signal would otherwise overlap with the X-ray signal, such that the signal-to-noise ratio would be too low for the shortest ToF (i.e. for the highest energy neutrons). During this experiment, 15 cm thick lead bricks were used in the front face of the scintillators and 5 cm on the top, bottom and rear faces. This allowed us to obtain clear neutron signals with the scintillator #16, for example, as presented in Fig. 11.

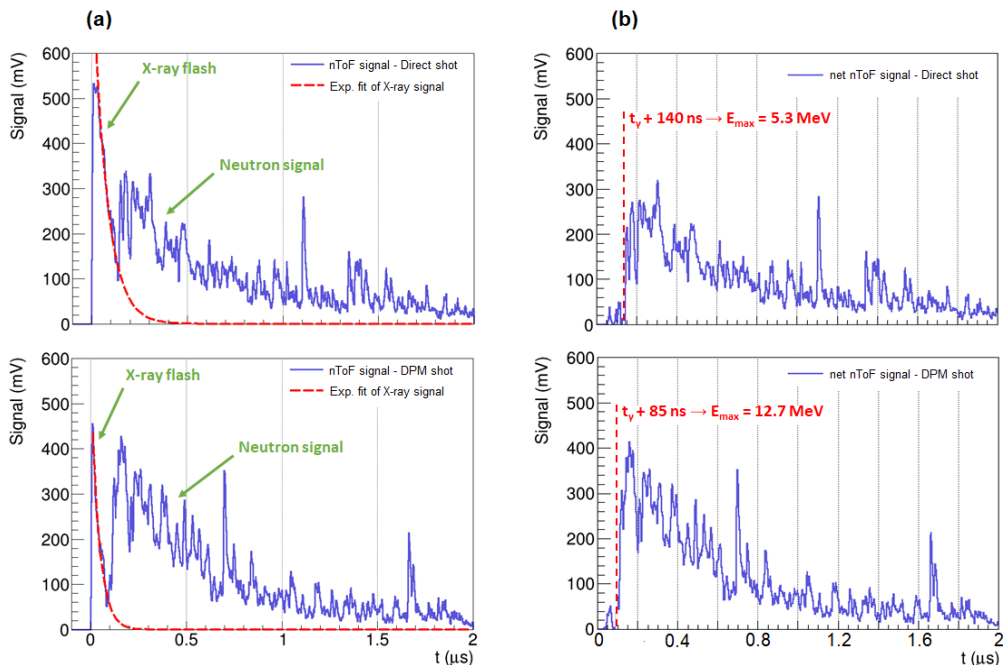

**Fig. 11.** (a) nToF signals obtained with the scintillator #16 placed at 4.95 m from the TCC and at 19° from the normal axis. These signals were acquired during a direct shot with a 1.5 µm Al target (top) and a DPM shot with a 200 nm Si + 50 nm Al target (bottom). The exponential fits of the X-ray signals used for the background noise subtraction are shown in red dashed lines. (b) Net neutron signals obtained after removal of the signal induced by the X-rays. The red dashed lines correspond to the ToF of the most energetic neutrons detected.

Fig. 11(a) shows similar amplitude for the neutron signals but significant discrepancies in X-ray emissions for shots with and without DPM. The integral of the X-ray flashes was calculated to quantify the charge induced by the X-rays (considering the measured voltage over time on a 50 Ω load). This leads to a factor of around 2.5 between the direct shot and the DPM shot, with 924.6 pC and 378.9 pC, respectively. Radio-photo-luminescence (RPL) dosimeters, placed at 1.15 m from the target chamber center and on the same axis as the scintillator #16, confirm this factor by measuring, respectively, doses of 16.1 µGy/shot and 5.8 µGy/shot during the series of direct and DPM shots. The DPM clearly affects the X-ray production, due to the possibility to shoot on thinner target inducing less X-rays. This allows to have a better signal-to-noise ratio for the short ToF and to measure more energetic neutrons, as shown in Fig. 11(b) where neutrons up to 12.7 MeV were detected.

To isolate the neutron signal, the X-ray-induced signals can be removed by subtracting exponential fit functions, which well describe the shape of the X-ray flashes (see Fig. 11(a)). The results of this background noise removal are presented in Fig. 11(b). The neutron signal obtained during the DPM shot presents a higher amplitude than that measured during the direct shot, especially for ToF from 85 to 500 ns, which corresponds to greater emissions of fast neutrons, from 500 keV to 12.7 MeV. As shown in Fig. 2, the proton spectra generated during the DPM shots have a greater proportion of high energy protons, which explains the more important emission of fast neutrons.

Thus, this diagnostic confirms the possibility to produce similar neutron emissions using the DPM, which is consistent with the results obtained via the bubble detectors and the activation samples, while emitting less X-rays.

Geant4 simulations of the nToF signals were made considering the simulated neutron spectra on the axis of the scintillator #16, for direct and DPM shots. A comparison between the measured and simulated results, normalized considering to the total number of neutrons observed from the $^{7}$Be measurement, are shown in Fig. 12 for both direct and DPM shots.

The experimental signal during the direct shot is much lower than expected at low ToF, a possible explanation being the detection limit induced by the important X-ray emissions which impairs the detection of high energy neutrons. Another explanation could be a fewer number of high energy protons generated during this shot compared to the spectrum considered in the simulations, inducing fewer high energy neutrons. Significant discrepancies also appear when the ToF increases, leading to an underestimation of low energy neutron emissions and/or an insufficient consideration of the scattered neutrons in the simulations. This observation can also be done on the comparison of the experimental and the simulated signals for the DPM shot. However, this comparison shows a very good agreement for low ToF (from 85 to 225 ns, which corresponds to neutrons between 2.2 and 12.7 MeV), demonstrating the importance of the DPM of reducing X-rays to improve our ability to detect high-energy neutrons.

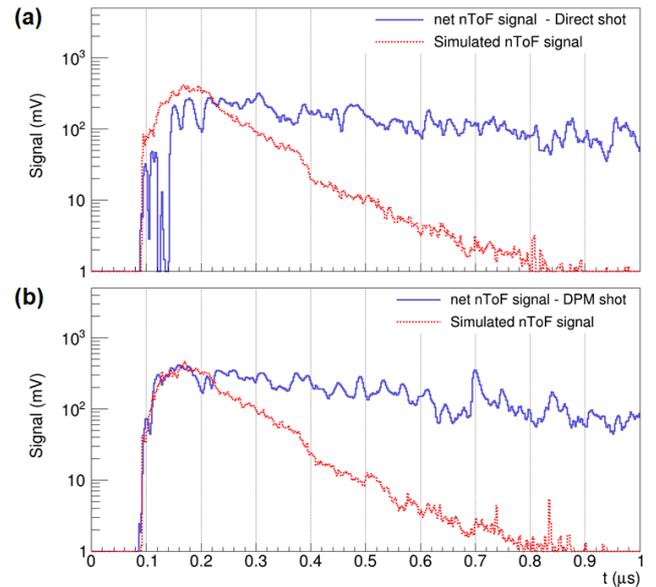

**Fig. 12.** Comparison between simulated (red) and experimental nToF signals (blue) for the direct shot (a) and the DPM shot (b).

Finally, an iterative algorithm using the simulated response function obtained with Geant4 simulations (see Section III.C) was developed to retrieve neutron energy spectra from the net nToF signals. This deconvolution procedure first calculates the number of neutrons, $N$, detected for each ToF, noted $t$, using the following equation:

$$N(t) = \frac{P(t)}{C(t)} \qquad (10)$$

Where $P(t)$, the number of photons for the bin at $t$ µs calculated from the measured voltage according to the energy calibration made with a $^{137}$Cs source. And $C(t)$, the efficiency calibration computed with the Geant4 simulations, giving a number of neutrons per photon (depending on the energy of neutrons considered and so on the ToF $t$).

But, as shown in Fig. 8, neutrons of a given energy also contribute to the signal at higher ToF (i.e. at lower energies). The number of neutrons $N(t)$ previously obtained should thus be subtracted by the contribution of higher energy neutrons (and lower ToF) to the signal. These contributions are considered as decreasing exponential functions with a slope depending on the neutron energy and so, on the ToF. The final number of neutrons at a given ToF $t$ can therefore be defined as:

$$N(t) = N(t) - \left( \sum_{i}^{n} N(t-i) \times e^{-A(t-i) \ast i} \right) \qquad (11)$$

Where $A(t-i)$, the slope of the exponential function describing the contribution to the signal of neutrons having interacted at $(t-i)$, and $i$ the bin width (2 ns, here).

Thus, neutron spectra can be extracted from the net nToF signals presented in Fig. 11(b). Fig. 13 shows a comparison between the simulated spectra, normalized by the number of $^7$Be nuclei measured, and the experimental nToF spectra obtained for the direct and the DPM shot. For the direct shot, the nToF spectrum is lower than the simulated one for energies above 2.5 MeV and higher below, which is consistent with the discrepancies observed in Fig. 12(a). A good agreement is however observed between the simulated and nToF spectra for the DPM shot, especially between 2.2 and 8 MeV. This good agreement is also consistent with the similar nToF signals obtained experimentally and from the simulation (Fig. 12(b)) in this energy range, supporting the nToF signal deconvolution algorithm. Discrepancies appear for energies above 8 MeV, this could be due to fewer number of high energy protons generated, inducing fewer high energy neutrons.

Both nToF spectra show much more emissions of low energy neutrons compared to the simulations. The integral of the nToF spectra, considering the large amount of low energy neutrons obtained, gives values of one order of magnitude higher than the neutron fluences measured with the bubble dosimeters in the same energy range. This low energy part seems therefore to be overestimated by the nToF diagnostic, which is probably due to the contribution of the scattered neutrons inside the experimental room, whose the probability of interaction with the detector can be significant, considering the dimensions of the scintillators.

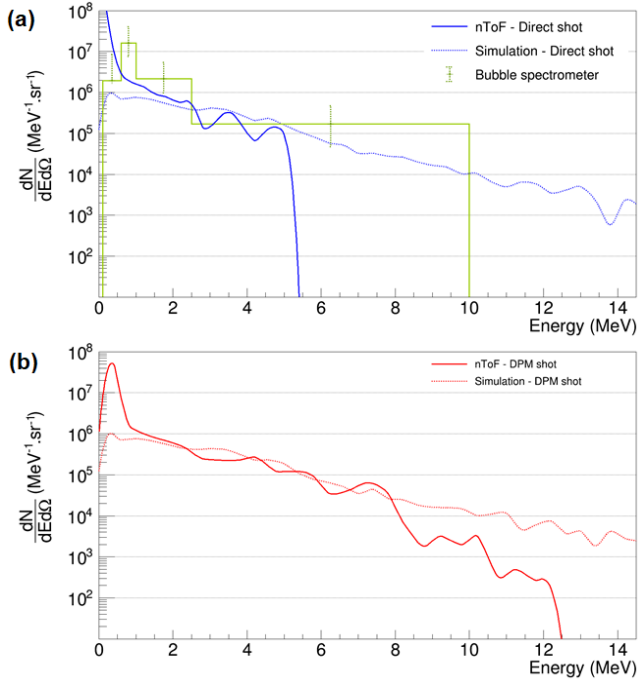

**Fig. 13.** Comparison between simulated and experimental neutron spectra measured using the nToF diagnostic during direct (a) and DPM (b) shots. The neutron spectrum obtained with the bubble spectrometer during the series of direct shots is shown on (a).

Finally, the normalization of the simulated spectra, by the number of $^7$Be measured, induces an overall good agreement between these simulated spectra and the nToF spectra, confirming the previous estimation of $3.93\times10^7$ neutrons emitted per shot, obtained from the measurement of the LiF converter by gamma spectrometry.

## V. SUMMARY & CONCLUSION

We have performed a detailed characterization of the neutron fields produced by the Apollon femtosecond laser, both with its inherent temporal contrast and with a contrast enhancement achieved using a DPM. Simulations were carried out using Geant4 to predict the characteristics of the neutron emissions, to design and calculate the expected activation of the samples and the response of the scintillators.

The neutron emissions were characterized using several detectors: activation samples, bubble detectors and scintillators. An estimation of the total number of neutrons emitted was obtained by direct measurement of the LiF converter by gamma spectrometry, leading to an average of $3.93\times10^7$ neutrons per shot, which is greater than the simulation predictions by a factor of approximately 1.4.

Neutron fluences of $1.689\times10^6$ ± 22.1% neutrons/sr/shot for direct shots and $1.136\times10^6$ ± 45.1% neutrons/sr/shot for DPM shots were measured between 2 and 10 MeV with the indium samples. The extended sensitivity range of the bubble dosimeters, from 100 keV to 20 MeV, pushed these values to $4.377\times10^6$ ± 17.4% neutrons/sr/shot for direct shots and $4.722\times10^6$ ± 40.8% neutrons/sr/shot for DPM shots.

A bubble spectrometer was used to determine the neutron energy distribution. Its result was found to be consistent with the simulation results, considering the total number of neutrons measured, especially for high energy neutrons. Significant differences appeared for low energy neutrons (< 2 MeV), where the simulations seem to underestimate the emissions. The same observation was also made with the results obtained from the nToF diagnostic. An iterative deconvolution algorithm was applied to the nToF signal, leading to the determination of neutron energy spectra, and showing good agreement with the simulations above 2 MeV.

However, these experimental spectra also present important emissions of low energy neutrons that was not expected by the simulations. This could be due to an insufficient consideration of the scattered neutrons in the simulations. Another explanation could lie in the proton spectra used to simulate the neutron emissions. The typical TNSA proton spectra are Maxwellian distributions usually described as a sum of two exponential functions with a slope at low energy greater than the one at high energy [51], while we used proton spectra considered as simple exponential functions. This consideration therefore tends to underestimate the number of low-energy protons and so the number of low-energy neutrons. An in-depth study of the origin of these significant emissions of low energy neutrons should be carried out with additional simulations or experimentally studied using the shadow cone technique [52,53].

Then, fewer X-ray emissions were noticed during the shots with a DPM according to the results obtained with the RPL X-ray dosimeters and the nToF detectors, which improves our ability to detect high energy neutrons with this diagnostic. Indeed, the improvement of the laser/target interaction, by reducing the prepulses, allows to shoot on thinner targets which produce less X-rays. At the same time, the results obtained by all diagnostics demonstrated that the production of neutrons was similar to that in direct shots, although the DPM reduces the on-target energy by 48%. Thus, the DPM seems to be an interesting tool to adjust the neutrons/X-rays ratio using different target thicknesses, which is a feature that could be used for some applications like combined neutron/X-ray radiography, allowing to probe both light and heavy materials.

Finally, the maximum laser energy available at Apollon will increase from 10 to 150 J with the progressive commissioning of the main laser beam (F1), inducing interesting prospects in the improvement of the proton acceleration and the neutron generation and making this facility a new potential neutron source which could be used for applications in a near future. Larger neutron emissions would also make it possible to activate more activation samples and therefore consider the possibility of carrying out neutron activation spectrometry.

## VI. ACKNOWLEDGMENTS

The authors acknowledge the national research infrastructure Apollon and the LULI for their technical assistance. The authors also thank Florian Negoita and Marius Gugiu for the design of the nToF detectors. P. A. Söderström acknowledge the support of the Romanian Ministry of Research and Innovation under research contract PN 23 21 01 06. This work was supported by funding from the European Research Council (ERC) under the European Unions Horizon 2020 research and innovation program (Grant Agreement No. 787539, Project GENESIS), by CNRS through the MITI interdisciplinary programs and by IRSN through its exploratory research program.